\documentclass[pdflatex,sn-basic]{sn-jnl}

\usepackage{graphicx}%
\usepackage{multirow}%
\usepackage{amsmath,amssymb,amsfonts}%
\usepackage{amsthm}%
\usepackage{mathrsfs}%
\usepackage[title]{appendix}%
\usepackage{xcolor}%
\usepackage{textcomp}%
\usepackage{manyfoot}%
\usepackage{booktabs}%
\usepackage{algorithm}%
\usepackage{algorithmicx}%
\usepackage{algpseudocode}%
\usepackage{listings}%

\theoremstyle{thmstyleone}%
\theoremstyle{thmstyletwo}%

\theoremstyle{thmstylethree}%

\begin{document}

\title[Article Title]{Gray Anchoring: a New Computational Theory for Biological Color Constancy}


\author*[1,2]{\fnm{Kai-Fu} \sur{Yang}}\email{yangkf@uestc.edu.cn}
\author[3]{\fnm{Dajun} \sur{Xing}}\email{dajun\_xing@bnu.edu.cn}
\author*[1,2]{\fnm{Yong-Jie} \sur{Li}}\email{liyj@uestc.edu.cn}

\affil*[1]{\orgdiv{MOE Key Laboratory for Neuroinformation, School of Life Science and Technology}, \orgname{University of Electronic Science and Technology of China}, \orgaddress{\city{Chengdu}, \postcode{610054}, \state{Sichuan}, \country{China}}}

\affil[2]{\orgdiv{The Yangtze Delta Region Institute (Huzhou)}, \orgname{University of Electronic Science and Technology of China}, \orgaddress{\city{Huzhou}, \postcode{313001}, \state{Zhejiang}, \country{China}}}

\affil[3]{\orgdiv{State Key Laboratory of Cognitive Neuroscience and Learning \& IDG/McGovern Institute for Brain Research}, \orgname{Beijing Normal University}, \orgaddress{\city{Beijing}, \postcode{100875},\country{China}}}


\abstract{It is still challenging for computer vision to imitate human color perception, e.g., color constancy, which is a fundamental perceptual ability in humans to perceive, interpret and interact with their surroundings. Among others, the anchoring theory provides impressive insights for human lightness perception, yet the specific anchoring rules underlying color constancy have remained contentious for decades. In this work, we introduced a novel computational theory—gray-anchoring (GA) theory—to explain how the early stage of visual system contributes to color constancy and demonstrate how our GA rule applies to the chromatic domain by identifying gray surfaces within complex scenes. Furthermore, we also demonstrate the potential neural implementation of gray-anchoring by quantitatively analyzing the computational flows of concentric double-opponent (DO) cells in V1. The simulational results show that the concentric DO cells have the ability to identify gray surfaces within color-biased scenes and these gray surfaces can then be used by the higher-level cortices to easily estimate the illuminant. This finding offers not only a clear functional explanation of the concentric DO receptive fields of this cell type in the visual system but also an effective and efficient solution to computational color constancy for computer vision.}


\maketitle

\section{Introduction}\label{intro}
Color is a fundamental aspect of our visual experience and plays a vital role in how we stably perceive, interpret and interact with the colorful world around us. However, how our brain perceives and interprets color is a complex process that is still not fully understood by scientists, which obviously hinders the field of computer vision to imitate human color perception. Among others, color constancy refers to the ability to perceive constant surface reflectance despite changes in illuminant, which is one of the important perceptual abilities in humans and other species \cite{dorr1996goldfish,werner1988color,garbers2015contextual}. This task is particularly challenging for computer vision systems because recovering surface reflectance from image luminance is an ill-posed problem \cite{gijsenij2011computational}, that is, the luminance values in the retinal image are determined by the product of the physical reflectance of surfaces and the intensity of the light illuminating those surfaces. In contrast, the human visual system achieves color constancy remarkably well in most natural situations. Despite significant efforts having been made to understand the computational principles underlying color constancy in early vision \cite{land1971lightness,gegenfurtner2003cortical,gao2013color}, how color constancy is achieved in the visual system is still not well-understood. 

To understand color constancy, it is essential to first address the more fundamental task of lightness constancy. Both tasks rely on a core computation that involves discounting the varying illuminant and estimating the reflectance of surfaces. Previous research has made numerous attempts to address the issue of lightness perception. For example, contrast theories \cite{helson1964adaptation, wallach1948brightness, gilchrist1996deeper} have posited that lightness constancy arises from the contrast interpretation, which is supported by the physiological evidence of lateral inhibition in biological visual systems \cite{baylor1971receptive,thoreson2012lateral}. Helson suggested that the perceived lightness of a surface is determined by comparing its luminance to the average luminance within the visual scene \cite{helson1964adaptation}. Additionally, Wallach introduced a simpler ratio theory of lightness, indicating that luminance ratios at edges are instrumental in explaining lightness constancy \cite{wallach1948brightness}.

However, the perceived relative luminance values, as interpreted through contrast theories, remain ambiguous. Specifically, the ratios of reflectances do not equate to the reflectances themselves. To address this issue, Gilchrist et al. proposed the famous anchoring theory attempting to explain how the visual system derives specific perceived lightness from relative luminance values extracted from a visual scene \cite{gilchrist1999anchoring, gilchrist1995anchoring, gilchrist1994anchoring}. Their anchoring theory states that absolute reflectances can only be recovered when mapping luminance to a standard scale with an anchor \cite{gilchrist1999anchoring}. 

According to the anchoring theory, finding suitable anchor points is critical for maintaining constant perception for lightness and color. Land and McCann made attempts to construct a computational theory of lightness perception within the human visual system, well-known as the Retinex theory \cite{land1971lightness}. Their groundbreaking work established a computational principle for determining the reflectance ratio between any two separated areas within a visual scene. Meanwhile, the Retinex theory suggests that surface lightness can be perceived under changing illuminant by anchoring the highest luminance in the visual scene to white, which is denoted as the highest-anchoring in this paper. Alternatively, the average-anchoring is also widely accepted, which takes the average luminance as a standard anchor for lightness perception. This average luminance rule is derived from Helson's theory \cite{helson1964adaptation} and is closely related to the famous gray world hypothesis in the chromatic domain \cite{buchsbaum1980spatial, hurlbert1986formal}.

\begin{figure*}[t!]
\centering
\includegraphics[width=\textwidth]{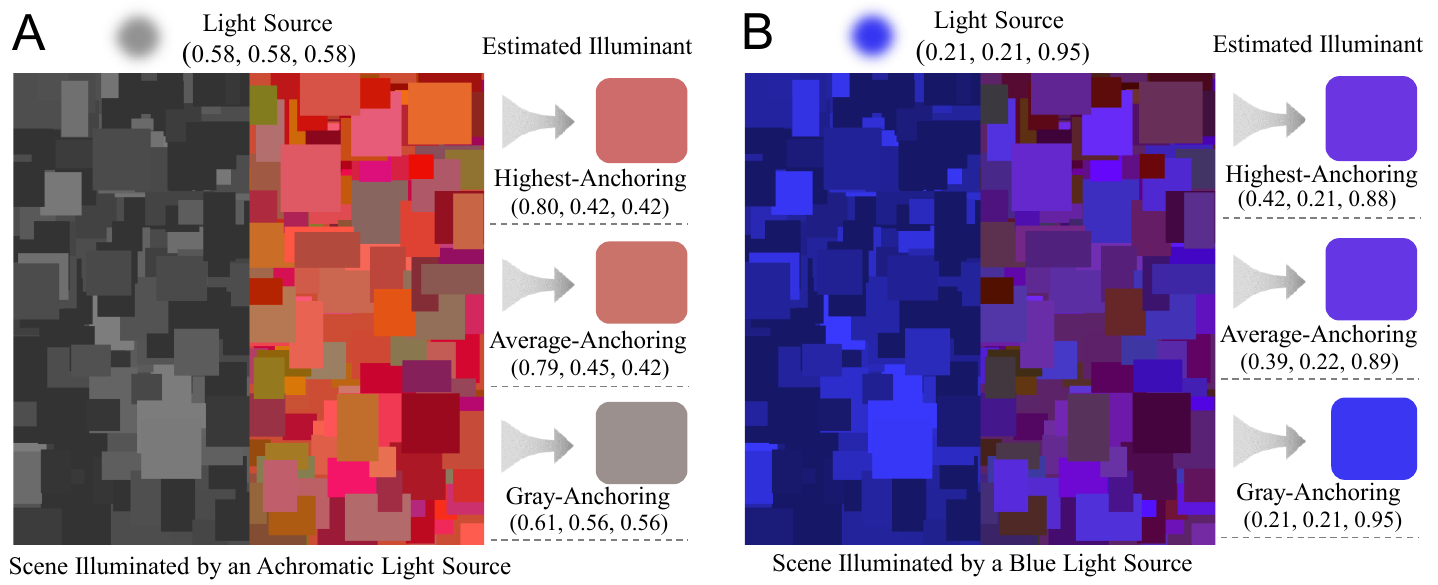}
\caption{The various anchoring rules for color constancy. (A) the scene is illuminated by an achromatic light source (indicated at the top of the image) and the estimated illuminants with different anchoring rules. (B) the scene is illuminated by a blue light source (indicated at the top of the image) and the estimated illuminants with different anchoring rules. The data in parentheses, formatted as $(r, g, b)$, represents the red, green, and blue component values of the light source or the estimated illuminants within  image space. The proposed gray-anchoring rule is implemented through the computational steps described in the section on Methods.}
\label{FigAnchoring}
\end{figure*}

Although the above two anchoring theories are widely understood in lightness constancy, how these anchoring rules apply to the chromatic domain (i.e., color constancy) remains underexplored. In the chromatic domain, the highest-anchoring and average-anchoring are computationally performed separately in each cone channel, i.e., long-, middle-, and short-wavelength cone channels, to remove the illuminant component \cite{land1977retinex, horn1974determining, buchsbaum1980spatial}. In the following text, to facilitate the expression of the computational process, the three channels are depicted as the red, green, and blue channels respectively within the context of describing image space. Obviously, this processing can achieve color constancy only when the reflectances of anchors in the three channels are equal with each other. Otherwise, anchoring the highest or average luminance values in each channel to white or gray will introduce additional color shifts in the recovered reflectance and fail to achieve color constancy \cite{jobson1997multiscale, barnard1998investigations}. As shown in Fig. \ref{FigAnchoring}, consider a scene in the Mondrian style that is composed of multiple surfaces with randomly varying sizes (Fig. \ref{FigAnchoring}A). Specifically, the left half of the scene is composed exclusively of gray surfaces, while the right half is composed of reddish surfaces. When the scene is illuminated by an achromatic light source, applying the highest-anchoring or average-anchoring rules results in the estimated illuminant to appear reddish. Similarly, when the scene is illuminated by a blue light source (Fig. \ref{FigAnchoring}B), the highest-anchoring or average-anchoring rules again yield erroneous estimations of illuminant. This is because the highest reflectance in the red channel usually comes from reddish surfaces, using the highest reflectance as an anchor could lead to inaccurate estimation of illuminant. Meanwhile, the average reflectance is also likely to be skewed toward reddish values due to the presence of reddish surfaces on the right half, causing average-anchoring to fail.

Additionally, the neural computation of the anchoring theory in the chromatic domain remains an unresolved issue. Land and McCann established the computational principle for determining the reflectance ratio between any two areas within a visual scene and emphasized the significance of edges as crucial sources of information for recovering reflectances under varying illuminant conditions \cite{land1971lightness}. The general neural implementation of obtaining the reflectance ratio is directly supported by the physiological evidences of lateral inhibition and center-surround receptive fields in biological visual system \cite{kuffler1953discharge, land1986alternative}. However, for color constancy, anchoring in the long-, middle-, and short-wavelength cone channels independently will lead to one of the main arguments arising from neurophysiological observations, that is, neural processing of color vision seems to be based on the color-opponency mechanism in the early visual system \cite{de1966analysis}, rather than independent processing in three cone channels. There are some valuable insights regarding the neural implementation of anchoring and color constancy in early visual system, but certain aspects remain unexplained \cite{conway2001spatial,shapley2019cortical}.

\section{Methods}
\subsection{The Gray-Anchoring Theory}
In this work, we propose a new gray-anchoring rule in the chromatic domain for color constancy. When anchoring to the intrinsic gray surfaces (these surfaces would appear gray if illuminated by a white light source), instead of the highest or average luminance values, the estimated illuminant according to the gray-anchoring rule will closely approximate the actual illuminant and align more consistently with human perception, as shown in Fig. \ref{FigAnchoring}.

\begin{figure*}[t!]
\centering
\includegraphics[width=\textwidth]{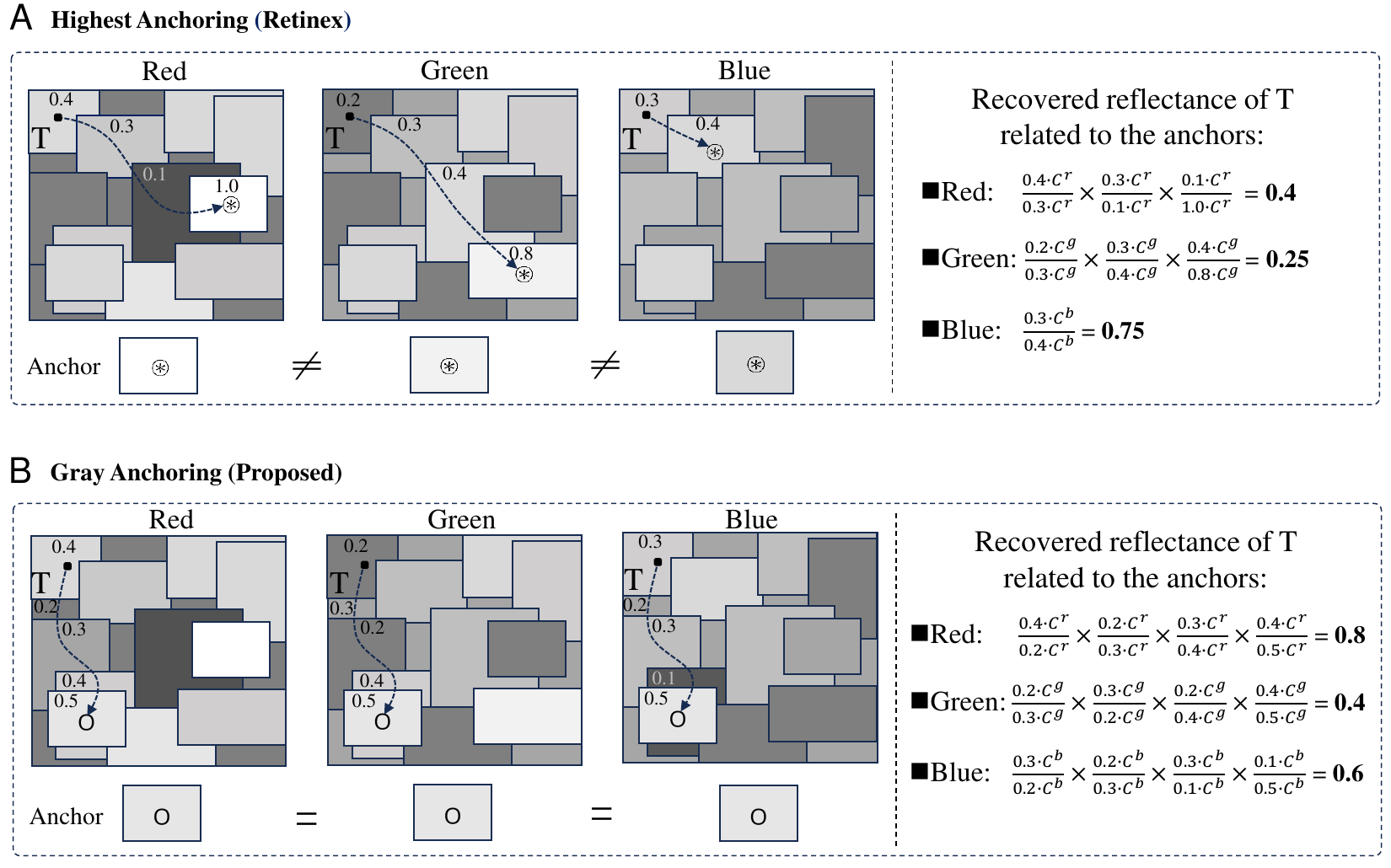}
\caption{The highest-anchoring rule and the proposed gray-anchoring rule in chromatic domain. (A) the highest-anchoring, $\circledast$ indicates the patches with the highest reflectance in the three color channels, (B) the gray-anchoring, $o$ indecates the patches with the equal reflectance in the three color channels. Note that the percentages on Mondrian-like patches refer to the reflectances of corresponding patches. $c^r$, $c^g$, and $c^b$ indicate the illuminant components of each color channel, which is assumed to be locally uniform. }
\label{FigGA}
\end{figure*}

In fact, from the perspective of computational implementation, we can also find that the highest- or average-anchoring rules encounter challenges in color vision. For example, in order to recover reflectance, Retinex establishes the standard reference as an area with the highest reflectance, and then performs a sequential product for any area related to that standard reference \cite{land1971lightness}. That means the Retinex method achieves surface reflectance recovery with  highest-anchoring rule following the anchoring theory proposed by Gilchrist et al. \cite{gilchrist1999anchoring}. Fig. \ref{FigGA}A illustrates the sequential product with highest-anchoring rule. Calculating the ratio between two adjacent patches can effectively eliminate the impact of illuminant. When identifying an anchor, the patch with the highest reflectance, the ratio of reflectances between any two widely separated areas can be determined by the sequential product of the luminance ratios along a path connecting two considered areas. Consequently, the reflectance of any area can be recovered using this sequential product approach, assuming the anchor’s reflectance is 1.0 \cite{land1971lightness}.

However, in the context of color vision, highest-anchoring determines reflectances in three independent cone channels: long-, middle- and short-wavelength cones, illustrated as the Red, Green and Blue channels in image space in Fig. \ref{FigGA}A. Unfortunately, such approach can lead to color shifts when the highest reflectances in these three channels are not identical (i.e., when the anchors are not achromatic or gray), shown in Fig. \ref{FigAnchoring}. Such discrepancies frequently occur in the real world. Consider a scene filled with green grass. The maximum references in three color channels are evidently unequal to each other. That means the anchors are green (rather than gray), which will result in inaccurate illuminant estimation. Fig. \ref{FigGA}A illustrates this issue, the recovered reflectances for the target area (denoted as 'T' in the figure)  may be imprecise because the anchor is not gray. Specifically, the sequential product in the three color channels is not related to the same standard. For instance, while the true references of the target area (T) in three channels are  0.4, 0.2, and 0.3, the recovered references turn out to be  0.4, 0.25, and 0.75.  

In the term of computational theory, we believe that gray surfaces play a crucial role in recovering reflectances under varying illuminant \cite{yang2015efficient}. This insight prompted us to revisit the anchoring theory and address the challenge encountered by highest or average anchoring. Our perspective is that achieving color constancy is feasible if the reflectances of the anchors are equal across all three color channels. Consequently, the references for the target area (denoted as ‘T’) can be accurately recovered using a sequential product approach based on the gray anchors, e.g., 0.8, 0.4, and 0.6 in Fig. \ref{FigGA}B. While the recovered references may numerically differ from the true target area references (0.4, 0.2, and 0.3), they still maintain color constancy. 

As a result, we formally propose a new gray-anchoring theory for color constancy in early vision, that is, our visual system achieves color constancy by anchoring to intrinsic gray surfaces. This new theory is built upon two critical preconditions inherited from the Retinex theory: (1) uniform illuminant within a limited range and (2) the significance of edges in determining reflectances. On one hand, the reflectance ratio between two adjacent patches can be obtained only when they are under the same illuminant. i.e., requiring locally uniform illuminant conditions. On the other hand, the reflectance ratio is meaningful (not zero) only when an edge exists between two adjacent patches. 

So far, one important issue that remains unresolved is how the visual system identifies intrinsic gray anchors under color-biased illuminant. This motivates us to explore the potential neural computational mechanisms underlying gray anchoring. We have set our sights on the color opponency mechanism regarded serve for color encoding in biological visual systems and argued that the color opponency mechanism in the early vision is possibly the neural basis for implementing the gray-anchoring function. 

\subsection{Neural Basis for Gray Anchoring}
\label{sec.do}
\begin{figure}[t!]
\centering
\includegraphics[width=10 cm]{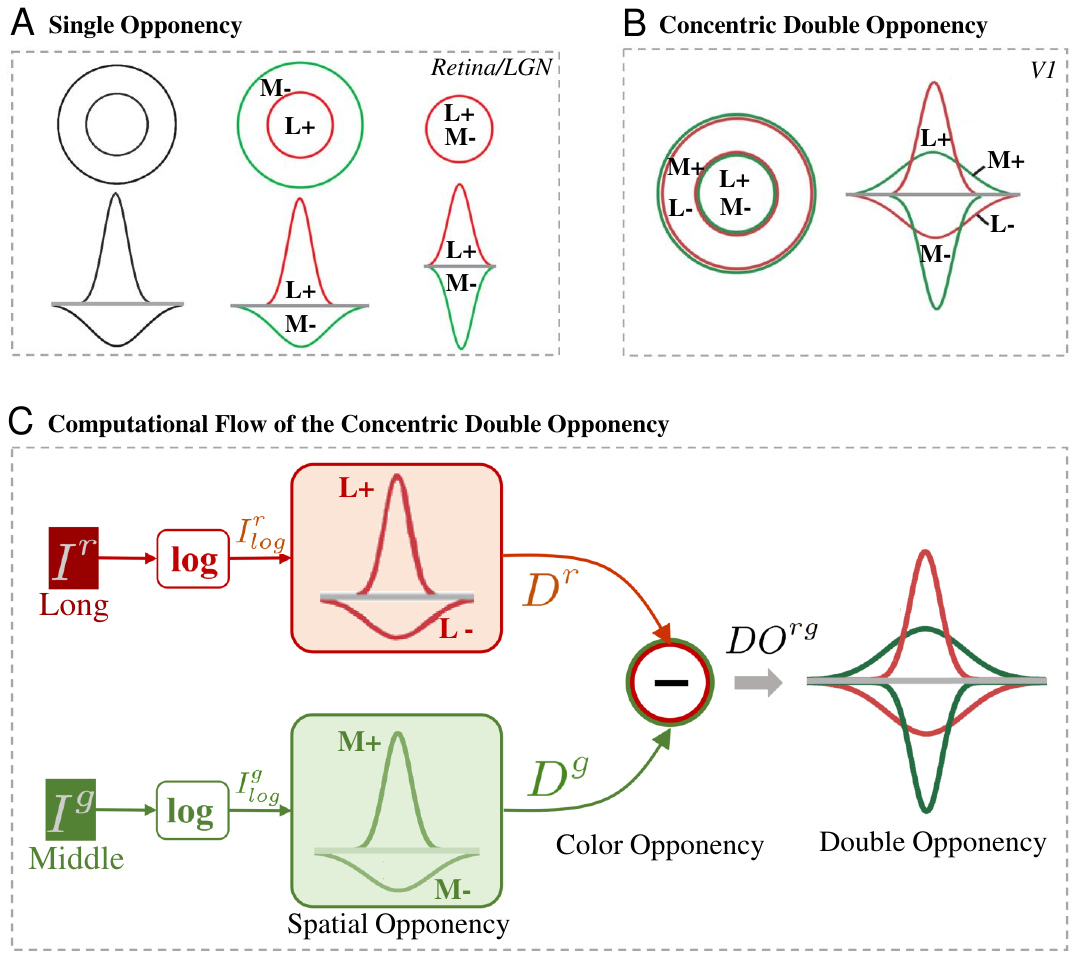}
\caption{Diagrams of typical color-opponent receptive fields in the early visual system. (A) The receptive field of typical single-opponent cells in Retina and LGN. (B) The receptive field of concentric DO cells in V1. (C) The computational flow of  the concentric double opponency, involving the input signals are initial processed by spatial opponent cells in each spectral band and then forwarded to color opponent cells.}
\label{FigCO}
\end{figure}

Although the color opponent theory was initially proposed by Hering in 1892 \cite{hering1964outlines}, it was not until 1965 that color opponent neurons were discovered in the primate visual system by DeValois et al. \cite{de1965analysis}. Typical single-opponent color neurons in retina and LGN have opponent inputs from different cone photoreptors, including four varieties: L-on/M-off, M-on/L-off, S-on/(L+M)-off, and (L+M)-on/S-off \cite{de1965analysis, conway2010advances, shapley2011color}. Additionally, the Type-II single-opponent color cells in the retina and LGN have also been reported to contribute to color vision \cite{conway2001spatial, wiesel1966spatial}. Plus the widely reported spatial-opponent neurons in the luminance stream \cite{kuffler1953discharge},  typical receptive fields of single-opponent cells in early vision have been shown in Fig. \ref{FigCO}A.

Afterwards, double-opponent (DO) cells, simultaneously performing color and spatial opponency, have been firstly found in the goldfish retina by Daw \cite{daw1967goldfish}.  Subsequently, such cells were also found in the striate cortex of non-human primates \cite{michael1978color1, michael1978color2, michael1978color3, de2021spatial, conway2001spatial}. Numerous studies have aimed to elucidate the functional roles of different types of DO cells \cite{conway2010advances, shapley2011color, de2021spatial}. Specifically, the concentric DO cells (Fig. \ref{FigCO}B), a type of color-sensitive cells found in the blobs of V1, remain enigmatic in terms of their functional role. These cells exhibit a center-surround receptive field organization: the central region is excited by one color and inhibited by its opponent color, while the surrounding region shows the reverse pattern. Some researchers believe that local computations by concentric DO neurons contribute to color constancy \cite{gegenfurtner2003cortical, shapley2011color, hurlbert2003colour}, but the specific mechanisms underlying their operation remain unclear.

To address the above-mentioned question, we analyze the computational flow to explain the organization of concentric DO receptive field. Theoretically, the concentric DO receptive field can be constructed with two potential computational flows, i.e., spatial opponency followed by color opponency and color opponency followed by spatial opponency. Considering both spatial and color opponent receptive fields are widely found in early vision, there is no reason to eliminate any of these two possibilities. However, according to our analysis, these two computational flows are computationally equivalent although with different computational steps. Therefore, we  discuss the first computational flow (shown in Fig. \ref{FigCO}C)) here and the details of alternative computational flow is demonstrated in the Supplementary Text and Supplementary Fig.S1.  

Following the Retinex point of view, the task of lightness perception is formally described as
\begin{equation}
\label{e1}
I^i = R^iC^i, i\in\{r,g,b,y\} 
\end{equation}
where $I^i$ is the reflected light (or the “image”) entering into the eyes, while $R^i$ is the spatial distribution of scene reflectances and $C^i$  denotes the spatial distribution of the source illuminant in the $i^{th}$ channel. We use $r$, $g$, and $b$ to indicate the corresponding signals from L-, M- and S-cones in retina, and $y$ represents the fused signal from L+M channel. 

The input signals are firstly transformed into logarithmic space in each spectral band to facilitate reflectance estimation and achieve color constancy \cite{fechner1966elements, land1971lightness}, i.e.,
\begin{equation}
\label{e2}
I_{log}^i = R_{log}^i+C_{log}^i
\end{equation}

Mathematically, a logarithmic function makes arithmetic division into a subtractive inhibition. As for the functional role of the logarithmic transform for the DO cells, Marr has strongly suggested that  “… the quantities R and G would have to be in logarithmic units. Such a cell would then act as capture detector of changes in color” \cite{marr2010vision}. Computational analysis in the following will further solid the necessity of logarithmic transform before spatial or color opponent computation.

According to the computational flow shown in Fig. \ref{FigCO}C, the input signals are firstly processed by the spatial opponent cells on each spectral band. For example, the on-center off-surround receptive fields will filter the signals as 
\begin{equation}
\label{e3}
\begin{split}
D_i &= I_{log}^i-F*I_{log}^i \\
&= R_{log}^i+C_{log}^i-F*[R_{log}^i+C_{log}^i] \\
&= R_{log}^i - \overline{R_{log}^i} + C_{log}^i - \overline{C_{log}^i} 
\end{split}
\end{equation}
where $F$ represents a Gaussian filter that characterizes the surround function, and $\ast$ denotes the convolution operation. Note that the center region of the receptive field is represented with one pixel of the image for convenience in formulaic expression. The bars (e.g., $\overline{R_{log}^i}$ and $\overline{C_{log}^i}$) indicate the spatially weighted average values. This center-surround operation is widely employed to describe the receptive fields of neurons in the early visual system \cite{kuffler1953discharge,rodieck1965quantitative}. According to Retinex theory, this operator can recover reflectance ratios by accounting for the smoothing of illuminant in a local region.
 
That means, as long as $ C_{log}^i\approx\overline{C_{log}^i}$ indicating illuminant within local region is smoothing or uniform, then
\begin{equation}
\label{e4}
D^i \approx  R_{log}^i - \overline{R_{log}^i}
\end{equation}

From Equation (\ref{e4}), we can find that $D^i $ is independent of locally smoothing illuminant ($ C_{log}^i\approx\overline{C_{log}^i}$). Meanwhile, the spatial opponent operator with center-surround receptive field detects differences in reflectance in the logarithmic space, i.e., reflectance ratio, which is meaningful only around the edges (i.e., $ R_{log}^i\neq \overline{R_{log}^i}$). This analysis highlights the crucial role of edges in reflectance recovery, as suggested by Land et al. in their Retinex theory \cite{land1971lightness}. 

Subsequently, the obtained reflectance ratios by spatial opponency are feedfwarded to concentric DO cells, whose receptive field can be constructed as illustrated in Fig. \ref{FigCO}C. The color opponent operation following spatial opponency can be written as 
\begin{equation}
\label{e5}
\begin{split}
DO^{rg} = D^r-D^g \\
DO^{by} = D^b-D^y
\end{split}
\end{equation}

Specifically, $DO^{rg}$ and $DO^{by}$ represent the responses of concentric DO cells with balanced cone inputs in the red-green and blue-yellow channels, respectively. Interesting insights emerge from Equation (\ref{e5}). When considering a local patch with gray reflectances (i.e., $R_{log}^r = R_{log}^g= R_{log}^b=R_{log}^y$), the responses of concentric DO cells should be zero due to $D^r=D^g$ and $D^b=D^y$, as indicated by Equation (\ref{e5}). In other words, gray surfaces can be identified by monitoring the response of concentric DO cells. The illuminant can be easily estimated and color constancy can be achieved referring to these gray surfaces, as described later.

It should be noted that $DO^{rg}\rightarrow 0$ and  $DO^{by}\rightarrow 0$ are necessary conditions for identifying gray surfaces. That is to say, there are inevitably certain non-gray (i.e., colored) surfaces that also meet these conditions. Furthermore, these discriminant conditions only apply in the local regions containing edges, i.e., non-smoothing in reflectance. That is because the smoothing reflectance will lead to the definition of reflectance ratio meaningless. For example, for the regions with smoothing reflectance, $D^i=0$ in Equation (\ref{e4}) will lead to $DO^{rg}=0$ or $DO^{by}=0$ in Equation (\ref{e5}), but we are unable to determine whether these areas are gray or not.

Consequently, we can conclude that the gray surfaces in a scene can be identified by the concentric DO cells in V1 and used to achieve some degree of color constancy.  Our computational analysis shows that color opponency is an efficient computational principle for detecting gray surfaces that can be used for achieving color constancy. Meanwhile, our computational model provides a qualitative description of the functional properties of concentric double-opponent cells in early vision that have puzzled visual scientists for many years \cite{gegenfurtner2003cortical, shapley2011color}. However, as indicated above, the discrimination condition for identifying gray surfaces based on concentric DO cells is only a necessary condition. Therefore, higher-level cortices and advanced cognitive mechanisms need to be involved for improving the robustness of identifying gray surfaces and achieving full color constancy \cite{witzel2018color, olkkonen2008color}.

\subsection{Brain-inspired Model for Color Constancy with Gray-Anchoring}
\label{sec.bimodel}
Computational flows shown in Fig. \ref{FigCO}C demonstrate the possible computational steps for the concentric DO cells with balanced cone inputs, but information represented in color opponent space primarily reflects color contrasts (e.g., red vs. green, blue vs. yellow), rather than the hues themselves. Therefore, one of puzzling issues is how the visual system obtains perceptual color experience (such as red, green, yellow, and blue)\cite{conway2023color} and achieves color constancy based on the opponency-based representation \cite{hurlbert2004color, foster2011color}. Additionally, the aforementioned computational analysis clearly demonstrates the rationality of concentric DO cells for identifying gray surfaces by finding their minimum response. This could raise another potential argument about the neural feasibility of computing the minimum response of DO cells for identifying gray surfaces, i.e., $DO^{rg}\rightarrow 0$ and  $DO^{by}\rightarrow 0$ in Equation (\ref{e5}). 

We contend that concentric DO cells, which receive balanced cone inputs in V1, encode information related to potential achromatic surfaces and subsequently contribute to achieving color constancy. The aforementioned issues can be elucidated when considering the involvement of higher visual cortices. In fact, numerous studies have demonstrated that the responses of color opponent cells are indeed transformed into a hue space within higher cortices for unique hue presentation \cite{kiper1997chromatic, stoughton2008neural,conway2007specialized,kim2020neural,li2022cone,liu2020hierarchical,conway2003colour,neitz2008colour}. For instance, neural representations within higher visual areas, such as V4, are posited to correspond to the perceived color \cite{kim2020neural}. Although the precise mechanism of the transition from color-opponent signals to color appearance remains unclear, there is a study that has suggested that neural circuits originating in V1 mediate an orderly information transition from cone-opponency to color appearance \cite{li2022cone}. Liu et al. have demonstrated a hierarchical hue representation within a uniform blob-like architecture along the visual hierarchy of macaque V1, V2, and V4 \cite{liu2020hierarchical}. Collectively, these studies endorse the plausibility of illuminant estimation occurring within higher visual cortices.

\begin{figure}[t!]
\centering
\includegraphics[width=12 cm]{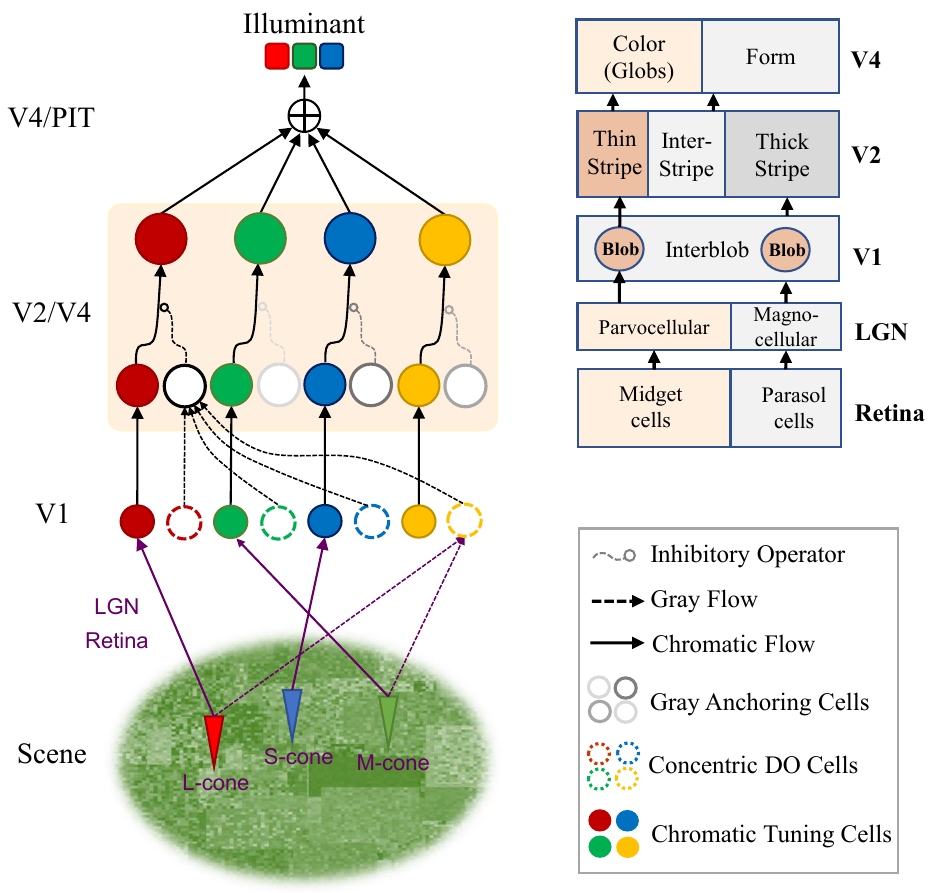}
\caption{A computational framework to account for illuminant estimation in the visual pathway. Chromatic-tuning cells in cortices encode unique hues along the visual hierarchy of macaque V1, V2, and V4. Meanwhile, the concentric DO cells with balanced cone inputs encode gray surfaces. The global scene illuminant can be represented in V4 or PIT by summarizing the hue responses of all surfaces that are firstly inhibited by the signals originating from the concentric DO cells with balanced cone inputs in V1. (Top-Right) The typical anatomical modular substructure of parallel pathways within the Retina-LGN-V1-V2-V4 hierarchy.}
\label{FigModel}
\end{figure}

We build a computational framework to quantitatively account for illuminant estimation along the visual pathway, as shown in Fig. \ref{FigModel}.  The model employs  a set of basic facts about the mechanisms of illuminant estimation based on double opponency. Firstly, chromatic-tuning cells in cortices encode unique hues along the visual hierarchy of macaque V1, V2, and V4, which could originate from color opponent cells with imbalanced cone inputs \cite{li2022cone, wu2024neural,xiao2003spatially}. Meanwhile, the concentric DO cells with balanced cone inputs also exist in V1 \cite{conway2010advances}, which can encode gray surfaces according to the computational flow illustrated in Fig. \ref{FigCO}C. Finally, the global scene illuminant could be represented in V4 or posterior inferior temporal cortex (PIT) by adequately described as a sum of hue responses of all surfaces \cite{bohon2016representation}. Specifically, the responses of hue-tuning V4 cells are proposed to be firstly inhibited by the signals originating from the concentric DO cells with balanced cone inputs in V1. Thus, the minimum response of concentric DO cells indicates weakest inhibition to the hue-tuning V4 cells, and hence gray surfaces will contribute more to global illuminant estimation. This is reasonable when considering that the apparent color of gray surfaces is solely due to the colored illuminant.

We further demonstrate the illuminant estimation in the visual system using the proposed computational framework.  With $DO^{rg}$ and $DO^{by}$ as the responses of concentric DO cells with balanced cone inputs (see Equation (\ref{e5})), the signals are then transferred from  V1 to V4, following the pathway that encodes gray information. Certain sub-groups of V2/V4 cells (hypothetical gray anchoring cells) can be assumed to encode the gray index of each surface by integrating the responses of two channels of color-opponency in V1, as

\begin{equation}
\label{e9}
GI =\sqrt{(DO^{rg})^2+(DO^{by})^2}
\end{equation}
Under this definition, a surface (or pixel) with lower $GI$ has higher probability of being gray.

On the other hand, hue-tuning V4 cells will encode the hue representation of each surface. To simplify the calculation, we utilize the reflected light (i.e., the input image $I^i, i\in\{r,g,b\} $) as the hue representation of each surface. Consequently, the scene illuminant can be described as a sum pooling of hue responses, which are modulated by inhibition from the concentric DO cells.
\begin{equation}
\label{e10}
e^i  = \sum\limits_{(u,v)} {\lfloor I^i(u,v) - w(u,v) \cdot I^i(u,v) \rfloor}, {\kern 1pt} {\kern 1pt} {\kern 1pt} {\kern 1pt} {\kern 1pt} i \in \{ r,g,b\} 
\end{equation}
where $\lfloor \cdot \rfloor$  denotes the floor function to ensure that the neuronal responses are non-negative, $w(u,v)$ represents the inhibitory weights at the spatial locations of $(u,v)$ based on the responses of concentric DO cells, which employ a commonly used sigmoid nonlinear activation function, i.e.,
\begin{equation}
\label{e11}
w(u,v)=A \cdot \big(\frac{1}{1+exp(-k \cdot GI(u,v))}-\tau\big) 
\end{equation}
where $A$ represents the gain factor, $k$ controls the sensitivity of selecting gray surfaces, and $\tau$ is a threshold. Specifically, we set $A=5$, $k=30$, and $\tau=0.5$ in the following experiments.

Finally, the estimated illuminant is used to correct the color-biased image and achieving color constancy with diagonal transform, i.e., von Kries Model \cite{kries1905influence}. The corrected image ($I_c^i,i\in\{r,g,b\}$) can be obtained by 
\begin{equation}
\begin{pmatrix}I_c^r\\I_c^g\\I_c^b\end{pmatrix}=\begin{pmatrix}1/e^r&0&0\\0&1/e^g&0\\0&0&1/e^b\end{pmatrix}\begin{pmatrix}I^r\\I^g\\I^b\end{pmatrix}
\end{equation}

We employed the angular error to evaluate the methods of illuminant estimation \cite{hordley2004re}, which is widely used in the field of computational color constancy and defined as
\begin{equation}
\epsilon = cos^{-1}\left( \frac{\textbf{I}_e \cdot \ \textbf{I}_g }{\|\textbf{I}_e \| \cdot  \ \| \textbf{I}_g \|} \right),
\end{equation}
where $\textbf{I}_e = [e^r, e^g, e^b]^T$ and $\textbf{I}_g$ are the estimated and ground-truth illuminants, respectively. $\| \cdot \|$ is the Euclidean norm of a vector.

\section{Results} 
\begin{figure*}[t!]
\centering
\includegraphics[width=\textwidth]{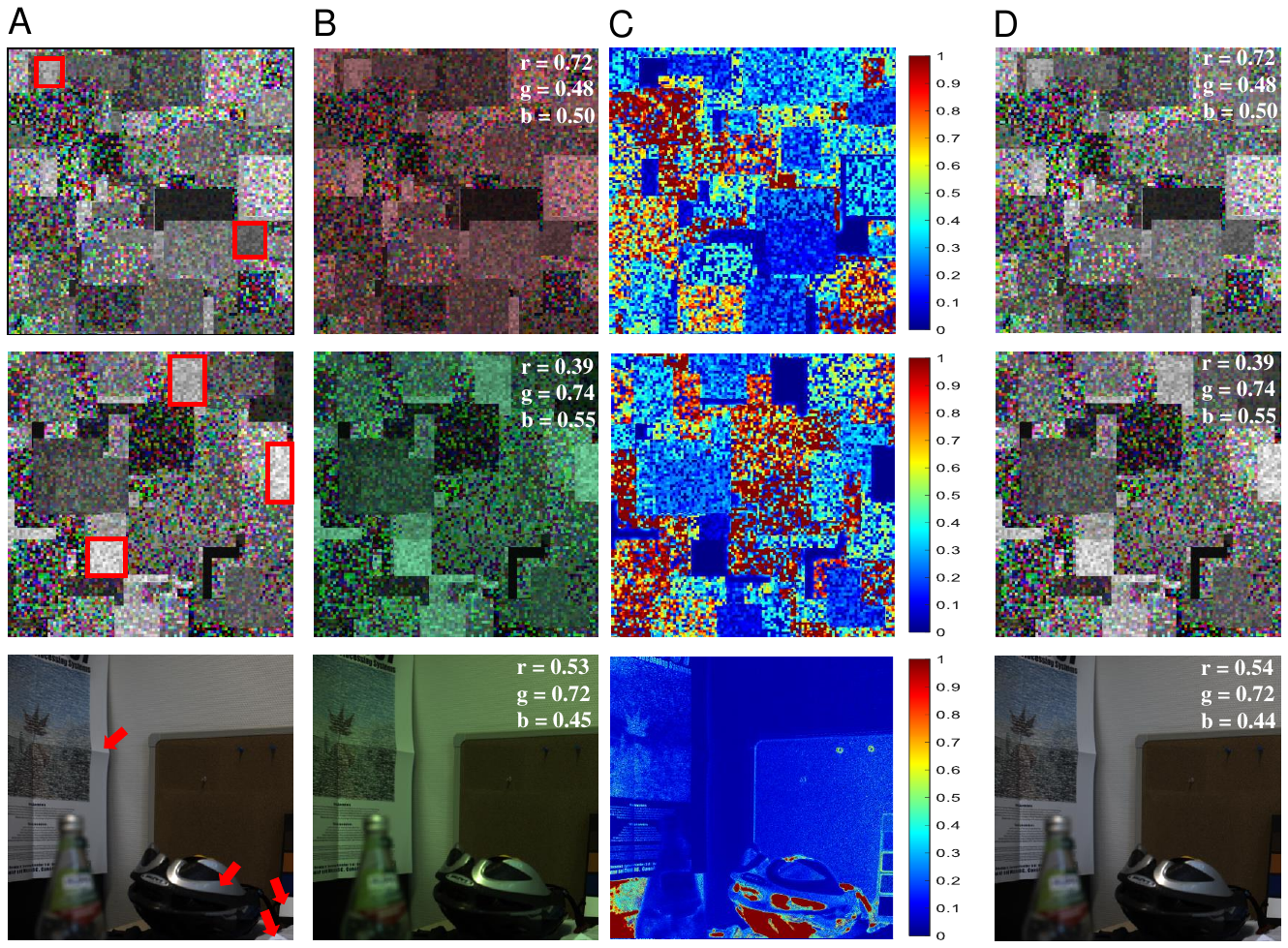}
\caption{Examples of two Meridian-like patterns with random speckled color noises (first two rows) and one natural image (last row) selected from ColorChecker\_REC dataset \cite{hemrit2019providing}. (A) canonical images, (B) color-biased images under global illuminant (noted on the top-right corner), (C) the responses of gray anchoring cells, and (D) corrected images by the estimated illuminant (noted on the top-right corner) based on the computational framework shown in Fig. \ref{FigModel}. Note that red rectangles and red arrows indicate the (approximate) gray surfaces in synthetic and real scenes.}
\label{FigRes}
\end{figure*}

We believe that one of the main functional roles of concentric DO cells is to encode gray information for color-biased scenes, which could significantly contribute to color constancy. Therefore, we verify the brain-inspired model in the computational color constancy. Fig. \ref{FigRes} illustrates three examples of color constancy on synthetic and real-world scenes. The synthetic images shown in Fig. \ref{FigRes} are generated as follows. A synthetic image is created by assembling multiple intensity patches of varying sizes, resulting in a Mondrian-like pattern. Furthermore, each patch contains random speckled color noise. Specifically, we set several patches to be gray, with equal pixel intensities across the red, green, and blue channels. As a result, we obtained the synthetic Mondrian-like images. Finally, we generate corresponding color-biased images by applying a global random illuminant to the Mondrian-like image, following the imaging model. 
Specifically, Fig. \ref{FigRes}A lists the canonical images, while Fig. \ref{FigRes}B presents color-biased scenes under global illuminant. The responses of concentric DO cells are provided in Fig. \ref{FigRes}C, which shows that minimum response consistently appears in areas with gray reflectances (indicated by red rectangles or arrows in Fig. \ref{FigRes}A). Finally, Fig. \ref{FigRes}D shows the well-corrected images by the estimated illuminant based on the computational framework shown in Fig. \ref{FigModel}.  More examples showing illuminant estimation with the proposed computational framework are listed in Supplementary Fig.S2.

\begin{figure}[t!]
\centering
\includegraphics[width=6cm]{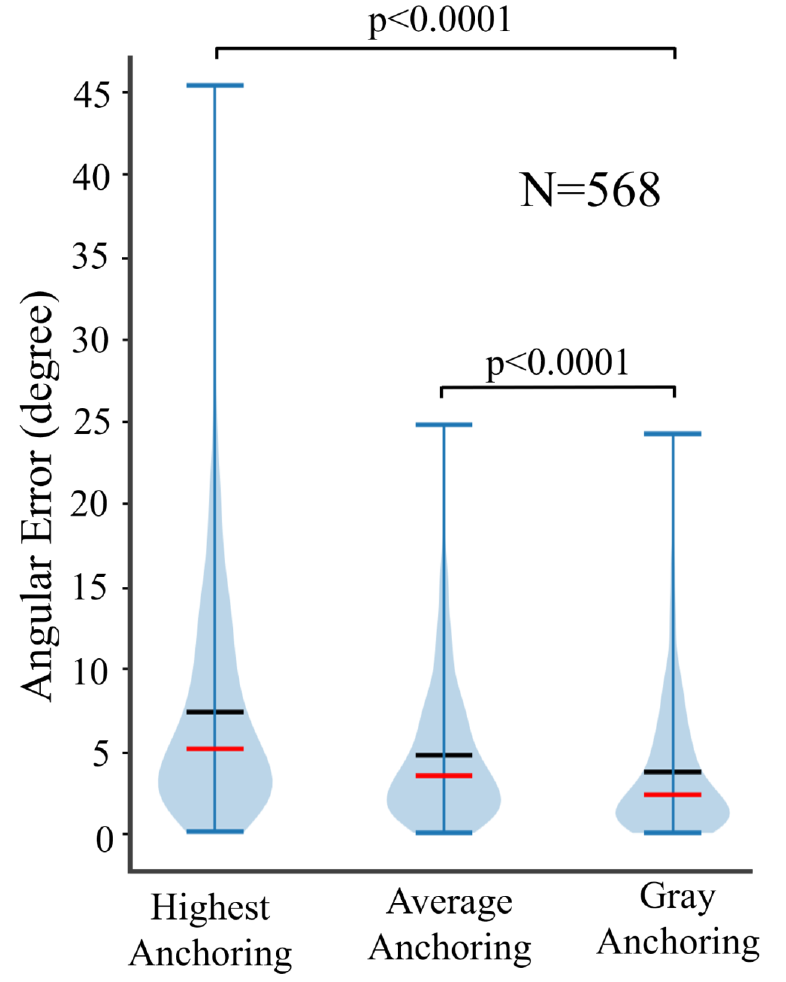}
\caption{Evaluating the performance of illuminant estimation with angular errors over the ColorChecker\_REC dataset \cite{hemrit2019providing} widely used in the field of computational color constancy. The number of images (N) is 568. The red lines indicate the median angular errors while the black lines indicate the mean angular errors. $p<0.0001$ indicates that there is a significant difference between the performances of the two compared methods using a two-sample t-test.}
\label{FigAE}
\end{figure}

On the other hand, the real image shown in Fig. \ref{FigRes} is sourced from the  widely-used ColorChecker\_REC dataset \cite{hemrit2019providing}. This dataset provides 568 high dynamic linear natural images with real illuminants for each scene, making it a widely used resource for evaluating color constancy methods in computer vision. Additionally, we further evaluated the performance of illuminant estimation methods on the total ColorChecker\_REC dataset. The distributions of angular errors between the estimated and true illuminant of images over the entire dataset are illustrated in Fig. \ref{FigAE}. Compared to classical illuminant estimation methods,  such as the highest-anchoring (i.e., White-Patch \cite{land1971lightness}) and the average-anchoring method (i.e., Gray-World \cite{buchsbaum1980spatial}), the proposed gray-anchoring method, which identifies gray surfaces using DO cells, achieved significant improvements.  For further details regarding the stimuli used in this experiment and the implementation for color constancy, please refer to the Methods section. Additional comparisons of illuminant estimation on the ColorChecker\_REC dataset are provided in Supplementary Fig.S3.

\section{Discussion}\label{sec12}
To achieve color constancy based on the anchoring theory, researchers have suggested that this process involves recovering reflectance under varying illuminants by comparing them to anchors that serve as references \cite{gilchrist1999anchoring, witzel2018color}. Computationally, the Retinex theory proposed reflectance recovery by implementing a sequential product for any area related to anchors \cite{land1971lightness}. Previous researchers have suggested using the space-average scene color (“gray-world” hypothesis) \cite{buchsbaum1980spatial, barnard2002comparison} or the color of the brightest surface in the scene (“bright-is-white” hypothesis) as the anchor for reflectance recovery \cite{land1971lightness, funt2010rehabilitation}. In computational color constancy, several methods focus on identifying intrinsic gray surfaces within images captured under colored light sources \cite{xiong2007automatic, li2010color, yang2015efficient}. These surfaces, which would appear gray if illuminated by a white light source, offer valuable cues for accurate light source estimation when precisely identified \cite{xiong2007automatic, li2010color}. However, these hypotheses lack support from neural computing in the visual system, and hence cannot explain the computational principles of color constancy in early vision. 

To explore the neural mechanisms of color constancy in V1, Gao et al. suggested a model following the computational flow of double opponency and suggested that the concentric DO cells can encode the illuminant information of the visual scene \cite{gao2013color,gao2015color}. Their study took a solid step forward in exposing the role of double opponency. But unfortunately, their work has not fully revealed the computational principle of color opponency in achieving color constancy. In this study, we present a novel perspective on the functional roles of concentric DO cells in V1, supported by qualitative analysis. Our findings also suggest a novel computational theory of color constancy in the early visual system, significantly extending the conventional anchoring theory of lightness perception. The proposed gray-anchoring theory serves as a bridge among Retinex theory, anchoring theory, and color-opponency theory in the context of color vision. Furthermore, our study offers a potential explanation for long-debated mechanisms underlying biological color constancy. Specifically, we propose that double-opponency in the early stage of vision serves as the neural basis for implementing the gray-anchoring function, contributing to color constancy. 

It is important to note that, in contrast to concentric DO cells, the more prevalent DO cells in V1 have oriented DO receptive fields that feature side-by-side spatially antagonistic regions with opponent cone weights \cite{johnson2001spatial, shapley2011color}. These oriented DO cells play a crucial role in color edge detection, with their orientation-selective receptive fields making this function both intuitive and understandable \cite{yang2013efficient, yang2015boundary}. Although this work focuses on concentric DO cells, oriented DO cells may also contribute to gray-anchoring processing by detecting colored edges. This is particularly relevant for scenes contain quite few detectable gray anchors, as suggested by \cite{yang2015efficient}. 

To what extent can color constancy be achieved through the gray-anchoring rule? This has been indirectly assessed in the field of computational color constancy. Our previous work introduced a color constancy framework (i.e., Gray-Pixel) by detection gray pixels in natural scenes \cite{yang2015efficient, cheng2024nighttime}, which can be considered as a specific implementation of computational flow suggested in the main text of this work. Subsequently, Qian et al. proposed a revised Grayness-Index algorithm \cite{qian2019finding}, which can be regarded as a specific implementation of another computational flow described in the supplementary material. Their validation results clearly demonstrate that the performances of these two methods, which anchor to gray pixels, are significantly superior to that of the highest-anchoring method (White-Patch) and the average-anchoring method (Gray-World). Specifically, on the INTEL-TAU dataset \cite{laakom2021intel}, Gray-Pixel and Grayness-Index achieved median angular errors of 2.2 and 2.3, respectively. In contrast, the White-Patch and Gray-World methods obtained median angular errors of 9.1 and 3.9, respectively. These studies suggested that detecting gray pixels is a promising way for computational color constancy. Additionally, the illuminant-invariant representation is also proved to contribute to subsequent tasks, e.g., object classification under varying illuminant \cite{funt2022laplacian}.

Furthermore, we consider double opponency a crucial factor in achieving color constancy, although it is not the entire story. Numerous studies have suggested that higher visual areas (e.g., V4) could be involved in generating the hue representation of color from the response of DO cells in V1 \cite{conway2007specialized, kim2020neural, li2022cone, liu2020hierarchical}. However, the neural mechanisms of hue perception are not yet fully understood \cite{kiper1997chromatic, stoughton2008neural, conway2007specialized,kim2020neural,li2022cone, liu2020hierarchical}. When combined with gray surfaces identified based on the response of DO cells, the hue representation of these gray surfaces could serve as an estimation of the illuminant. Hence, these studies support the feasibility of estimating illuminant in the visual system. However, according to the proposed brain-inspired model, two key hypotheses require further validation in visual neuroscience. First, we suggest the existence of a type of cell serving for gray anchoring in areas V2/V4, which would integrate responses from DO cells in V1 via an energy model as described in Equation (\ref{e9}). Additionally, the output of these gray anchoring cells is hypothesized to provide inhibitory feedforward input to chromatic tuning cells, thereby contributing to illuminant estimation by inhibitory all color surfaces. Therefore, the gray anchoring cells should respond to all color stimuli and indicate the saturation, without hue selectivity. 

A remaining question is how the visual system maintains color constancy in scenes that contain no achromatic pixels. Firstly, we validated the Gray-Pixel Hypothesis in previous work, demonstrating that most natural scenes contain some gray (or at least approximately gray) pixels \cite{yang2015efficient}. This environmental regularity provides a plausible evolutionary basis for the biological visual system to achieve color constancy by anchoring on achromatic cues. Moreover, higher-level cortices beyond the visual areas are also believed to contribute to color constancy in challenging scenes. That is, the color perception of some specific scenes (e.g., the scenes containing no gray pixels) could benefit from the higher cognitive feedback. For example, the memory color effect demonstrates that memory colors influence observers' perception on the colors of the objects \cite{witzel2018color, olkkonen2008color, hansen2006color}. The familiar colors of real objects in natural scenes may affect the perception of surface color \cite{hansen2006color, olkkonen2008color, foster2011color}. Identifying an object has a measurable effect on color perception, and this effect remains robust under varying illuminant \cite{olkkonen2008color}. This suggests that object recognition and memory color are additional mechanisms for color constancy.

\section*{Acknowledgements}
This study was supported by the STI2030-Major Projects (2022ZD0204600) and the National Natural Science Foundation of China (T2541042,2476050,32571280). This work was also partly supported by the Huzhou Science and Technology Program (2024GZ12) and the Fundamental Research Funds for the Central Universities (Y03023206100215).

\section*{Competing interests}
There are no competing interests to declare.

\section*{Correspondence}
Correspondence and requests for materials should be addressed to
Kai-Fu Yang or Yong-Jie Li.

\section*{Data availability}
the dataset (i.e., ColorChecker\_REC) used for evaluation is openly accessible at http://colorconstancy.com/evaluation/datasets/index.html.





\bibliography{mybib.bib}

\end{document}